\documentclass{iopjournal}

\usepackage{amsmath}
\usepackage{bm}
\usepackage{mathtools}
\usepackage{amssymb}
\newcommand \beq{\begin{equation}}
\newcommand \eeq{\end{equation}}

\newcommand{\Nc}{N_{\rm c}}

\newcommand{\lqcd}{\Lambda_{\rm QCD}}

\newcommand{\vq}{\vec{q}}
\newcommand{\vk}{\vec{k}}

\newcommand{\la}{\langle}
\newcommand{\ra}{\rangle}

\newcommand{\calB}{\mathcal{B}}

\newcommand{\calP}{\mathcal{P}}

\newcommand{\rmd}{\mathrm{d}}

\newcommand{\rme}{\mathrm{e}}

\begin{document}

\articletype{Article type} 

\title{A quarkyonic matter model}

\author{Toru Kojo \orcid{0000-0001-5656-3652} }

\affil{Theory Center, IPNS, High Energy Accelerator Research Organization (KEK), 1-1 Oho, Tsukuba, Ibaraki 305-0801, Japan}

\affil{ Graduate Institute for Advanced Studies, SOKENDAI, 1-1 Oho, Tsukuba, Ibaraki, 305-0801, Japan }


\email{torukojo@post.kek.jp}


\begin{abstract}
Quarkyonic matter is a state of matter in dense QCD whose bulk thermodynamics is dominated by quarks, while low-energy excitations remain confined. 
This picture leads to a crossover description from baryonic matter to quark matter,
 which is triggered by the saturation of quark states in dense matter ({\it quark saturation}).
The crossover driven by the quark saturation accompanies rapid growth in pressure but moderate increase in energy density, 
resulting in a peak in the sound speed which has been indicated by observational constraints from neutron star physics.
The quark saturation can occur at a few times nuclear saturation density, which is smaller than the density at which the baryon cores of $\sim 0.5$--$0.8$ fm spatially overlap.
In this contribution we discuss an ideal model of quarkyonic matter, the IdylliQ model, and we explicitly describe how the baryon and quark occupation probabilities are related,
and explain how stiffening of matter occurs.
The model is further applied to charge neutral matter including hyperons,
and it is shown that the statistical constraints at quark level induce effective repulsion among different baryon species,
mitigating the hyperon softening problem in neutron star physics.
\end{abstract}

\section{Introduction} \label{sec:intro}

\begin{figure}[t]
\vspace{-1.cm}
\centering
\includegraphics[width=.7\textwidth]{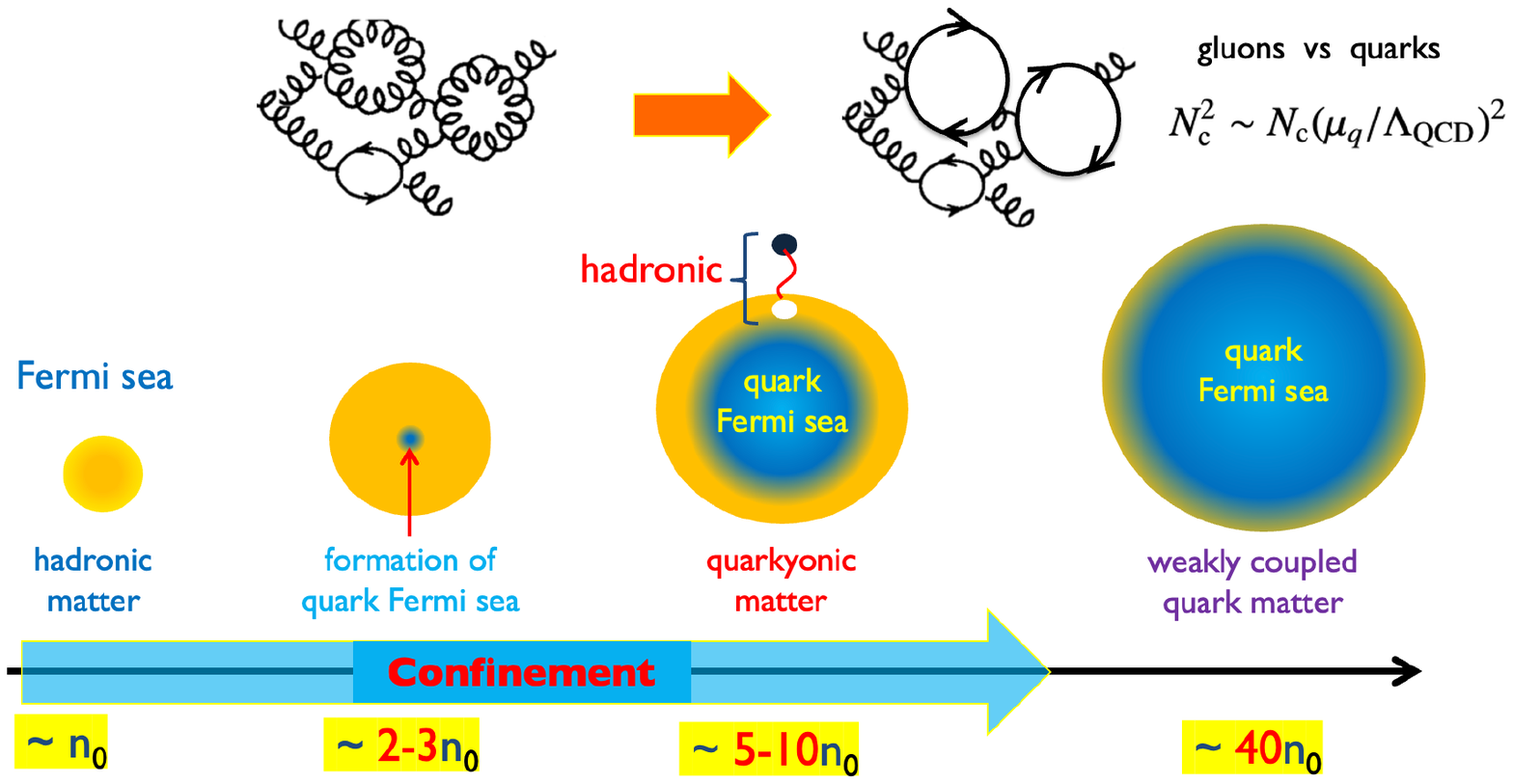}
\vspace{-1.cm}
\caption{ 
A quarkyonic matter scenario for the evolution of strongly interacting matter from the nuclear to the quark regime.
At large $\Nc$, gluonic contributions dominate over quark screening effects up to
$\mu_q \sim \Nc^{1/2}\Lambda_{\rm QCD}$, allowing confinement to persist at high density.
As a result, a quarkyonic regime emerges, characterized by a quark Fermi sea in the bulk
with confined, hadronic excitations near the Fermi surface.
Estimates for the real world with $\Nc=3$ are shown as a guideline for relevant densities.
}
\label{fig:quarkyonic}
\end{figure}

\begin{figure}[t]
\vspace{-1.5cm}
\centering
\includegraphics[width=.7\textwidth]{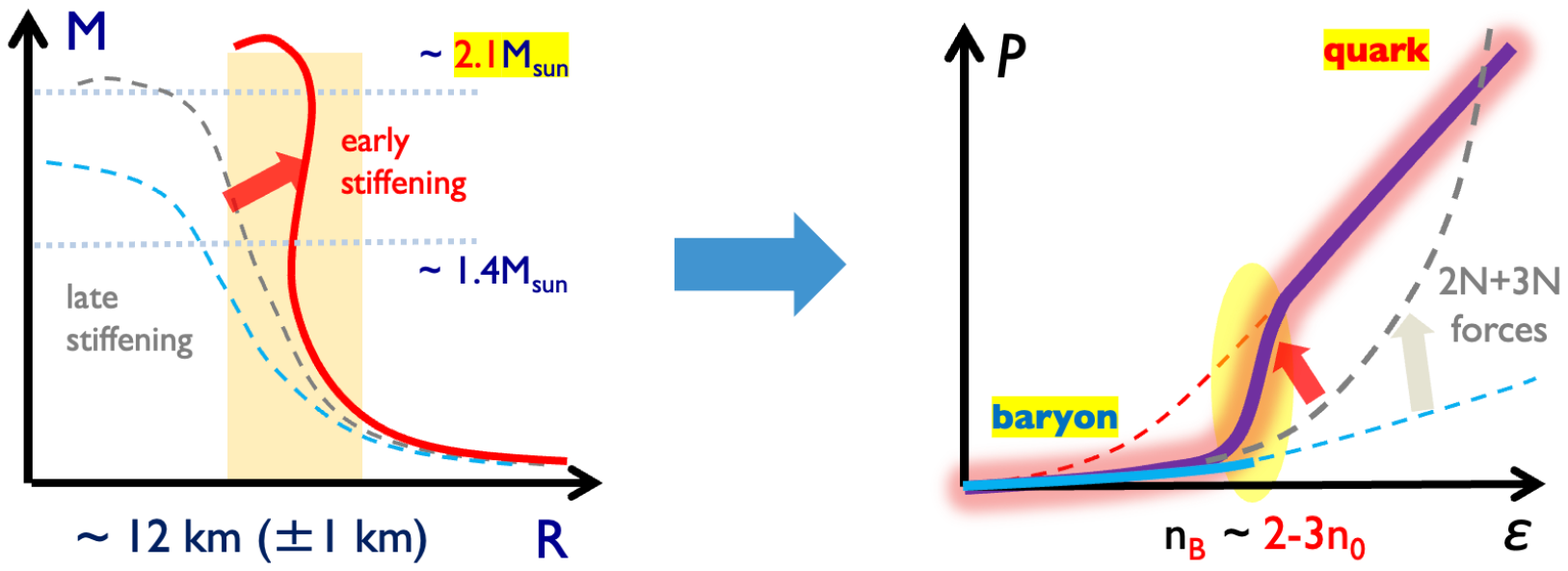}
\vspace{-1.5cm}
\caption{ 
Observational constraints on neutron star $M$-$R$ relations and the corresponding inference which suggests that
the EOS stiffens rapidly around 2-3$n_0$ ($n_0\simeq 0.16\,{\rm fm}^{-3}$: nuclear saturation density $\simeq$ nucleon density in typical nuclei) 
and approaches the quark matter behavior with $P \simeq \varepsilon/3$.
}
\label{fig:EOS_MR}
\end{figure}

Quarkyonic matter is a state of matter in which the bulk thermodynamics is dominated by quarks, while low-energy excitations remain confined. 
This hypothesis was proposed by McLerran and Pisarski in 2007 \cite{McLerran:2007qj}, through their attempts to establish the QCD phase diagram 
at a large number of colors, $\Nc \gg 1$ \cite{tHooft:1973alw,Witten:1979kh}.
At large $\Nc$, gluons, which are responsible for non-perturbative effects and confinement, are unaffected by in-medium screening due to quarks.
Even though whether the large $\Nc$ limit is close to the reality of $\Nc=3$ world remains a quantitative issue (see, e.g., Ref.~\cite{Shuster:1999tn}),
this setup at least sharpens our question toward confinement, the meaning of which in a dense medium would be somewhat obscured otherwise.

The quarkyonic matter hypothesis gives a natural microscopic baseline for a crossover transition \cite{Schafer:1998ef,Hatsuda:2006ps,Baym:2017whm}
from baryonic to quark matter (modulo possible first order transitions associated with pairing effects).
During this crossover transition, the effective degrees of freedom changes, depending on the domain of momentum space \cite{Fukushima:2020cmk}, see Fig.~\ref{fig:quarkyonic}.
As the baryon density ($n_B$) increases, 
in the bulk part of the baryon Fermi sea the natural degrees of freedom smoothly change from baryons to quarks,
while near the Fermi surface, baryonic composites survive as they do in vacuum.
As a result, we find a quark Fermi sea with a baryonic Fermi surface.
At sufficiently high density, the in-medium screening eventually cuts off non-perturbative gluons
and a quark matter approaches a conventional weakly coupled gas of quarks \cite{Freedman:1976ub,Kurkela:2009gj}.

Unlike conventional hybrid models with the first-order baryon-to-quark {\it phase} transitions,
the crossover transition drives a rapid stiffening of matter, 
i.e., the pressure ($\calP$) increases rapidly for moderate changes in the energy density ($\varepsilon$),
accompanying a peak in the sound speed, $c_s = (\partial \calP/\partial \varepsilon)^{1/2}$ \cite{Masuda:2012kf,Masuda:2012ed,Kojo:2014rca}.
Unlike the 1st order phase transitions,
the transition is triggered not because quark matter is energetically more favored than baryonic matter, 
but because the transition is {\it forced} to occur due to statistical constraints, i.e., the Pauli principle acting on quarks inside baryons \cite{Kojo:2021ugu}.
In models of quarkyonic matter, we closely pack many baryons into a dense regime,
and then we find that quarks saturate the available states, which we call the {\it quark saturation} \cite{Kojo:2021ugu}.
The saturated quark states inevitably lead to the formation of the quark Fermi sea.

The stiffening in the crossover occurs, according to model calculations with reasonable parameters,
at densities around 2--3$n_0$ ($n_0$: saturation density).
This is the domain where many-body repulsions become important in purely nuclear matter calculations,
and this density is about half the density at which baryons with a radius of $\sim 0.6$ fm spatially overlap. 

In the crossover description, the stiffening occurs more quickly than in typical nuclear models such as APR \cite{Akmal:1998cf} or ChEFT \cite{Drischler:2020fvz},
and this feature results in neutron star radii for massive neutron stars of $\sim 2M_\odot$ 
that are comparable to or even larger than those of $\sim 1.4M_\odot$.
The inference from NICER observations for PSR J0740+6620, the gravitational wave detection from GW170817, and nuclear physics constraints, 
suggests $R_{1.4} \sim R_{2.1} \sim 12$ km,
favoring a rapid stiffening around 1.5--3$n_0$ \cite{Drischler:2020fvz,Kojo:2021wax} (Fig.~\ref{fig:EOS_MR}).
While the NICER estimates contain systematic uncertainties,
the future detection of high frequency gravitational waves from post-mergers, especially $f_2$ modes, 
are expected to provide more information on the size of high mass neutron stars \cite{Huang:2022mqp,Hensh:2024onv}.

In this article, we present an ideal model of quarkyonic matter, the {\it IdylliQ} model \cite{Fujimoto:2023mzy}, to represent the key concepts of quarkyonic matter.
In Sec.~\ref{sec:schematic_nuclear_quark}, we review nuclear and quark equations of state in a schematic manner 
and explain why transitions to quark matter may stiffen equations of state. 
In Sec.~\ref{sec:q-sat}, we discuss the IdylliQ model and summarize the assumptions used.
In Sec.~\ref{sec:hyperons}, we extend the discussion to the multi-flavor cases, and argue how the quark saturation effects mitigate the hyperon puzzle.
Sec.~\ref{sec:summary} is devoted to summary.

\section{ Nuclear and quark matter equations of state: a schematic view } \label{sec:schematic_nuclear_quark}

In this section, we try to differentiate nuclear matter with many-body repulsion from quark matter.
In the case of the first order phase transition,
it is easy to characterize the boundary between nuclear and quark matter.
In the case of the crossover scenarios, however, we need more detailed discussions based on the properties of microphysics.

\subsection{ Nuclear matter and many-body repulsion } \label{sec:nuclear}

A schematic model for nucleonic matter (per flavor and spin) can be parametrized as 
\beq
\varepsilon_N (n_B) = m_N n_B + a \frac{\, n_B^{5/3} \,}{\, m_N \,} + b n_B^\alpha \,,
\eeq
where $m_N$ is the nucleon mass and we assume $a,b >0$.
The first, second, and third terms are the mass energy, kinetic energy, and the interaction energy from $\alpha$-body contact interactions.
The pressure can be computed using the thermodynamic relation $\calP = n_B^2 \partial (\varepsilon/n_B)/\partial n_B$,
\beq
\calP_N = \frac{\, 4 a n_B^{5/3} \,}{\, 3m_N \,} + b ( \alpha - 1 ) n_B^\alpha \,.
\eeq
It should be emphasized that, in the dilute regime, the energy density is dominated by the mass energy term, 
but it drops off from the pressure. The kinetic energy is small in both $\calP_N$ and $\varepsilon_N$.
Hence, unless the interactions play a crucial role, we find $\calP_N \ll \varepsilon_N \simeq m_N n_B$, meaning that the equation of state is soft.

Meanwhile, at very high density, the extrapolation of the above expressions
leads to the dominance of interactions for $\alpha >1$.
Asymptotically $\varepsilon_N \sim bn_B^\alpha$ and $\calP_N \sim b(\alpha-1) n_B^\alpha$,
and we find that the pressure-energy relation and the squared sound speed behaves as \cite{Kojo:2025vcq}
\beq
\calP_N \sim (\alpha-1) \varepsilon_N \,,~~~~~~~~c_s^2 \sim \alpha - 1 \,.
\eeq
For the two-body dominance ($\alpha=2$), the asymptotic behavior is $c_s^2 \sim 1$, close to the causal bound, $c_s^2 \le 1$.
The three-body dominance ($\alpha=3$) leads to $c_s^2 \sim 2$, violating the causal bound.
In the case of the famous APR \cite{Akmal:1998cf} or Togashi \cite{Togashi:2017mjp} equation of state, the causality is violated around $\sim 5.5n_0$ \cite{Baym:2019iky}.

To satisfy the $2M_\odot$ constraint on the neutron star mass,
usually two-body repulsion is insufficient so that three-body repulsions are employed.
However, there are several problems in models that rely heavily on three-body repulsions.
First, the importance of three-body repulsion raises a question concerning the convergence of many-body expansion.
Second, the violation of the causality at large $n_B$ suggests the necessity of more-body attractions which cancel the three-body repulsion;
it is difficult to construct systematic descriptions.
Last but not least,
the power-like growth of three-body forces, $\varepsilon_{\rm 3-body} \sim n_B^3$, is presumably too mild to explain $R_{1.4} \sim R_{2.1}$ \cite{Kojo:2021wax};
if we constrain the size of three-body forces to be consistent with nuclear matter properties at saturation,
the extrapolation of nuclear becomes sufficiently stiff around 3--5$n_0$ and then violates the causal bound around 5--6$n_0$.
This results in $R_{1.4} - R_{2.1} \simeq 1$ km which is close to the edge of or outside the NICER's band.
Although we need better quality of data with less systematic uncertainties to establish the last point,
the situation may be improved in the future detection of the high frequency gravitational waves
from post mergers whose $f_2$ mode is strongly correlated with the radii of high mass neutron stars \cite{Huang:2022mqp,Hensh:2024onv}.

\subsection{ Quark matter } \label{sec:quark}

We next turn to equation of state for quark matter.
To compare it with nuclear matter,
we first consider non-relativistic quark matter,
setting aside relativistic effects.
We parametrize it as (per flavors and spins) 
\beq
\varepsilon_Q = \Nc \bigg( m_Q n_B + a \frac{\, n_B^{5/3} \,}{m_Q} \bigg) + \varepsilon_Q^{\rm int}\,,
\eeq
where we wrote the factor $\Nc=3$ explicitly to count the same numbers of quarks for the nucleonic and quark matter cases.
We note that a quark at a given color appears in a single baryon, so $n_B = n_Q^R=n_Q^G=n_Q^B $.
Hence, the numerator of the kinetic term is $n_B^{5/3}$ as in the baryonic kinetic term. 
The prefactor $a$ is also common.
The corresponding pressure is
\beq
\calP_Q = \Nc \frac{\, 4 a n_B^{5/3} \,}{\, 3m_Q \,} 
+ n_B^2 \frac{\, \partial (\varepsilon^{\rm int}_Q/n_B) \,}{\, \partial n_B \,}
\,.
\eeq
The remarkable difference between nuclear and quark matter can be seen in the kinetic terms.
In nucleonic matter we found $\calP_N^{\rm kin} \sim n_B^{5/3}/ m_N \sim  n_B^{5/3}/\Nc m_Q$, 
while in quark matter we find $\calP_Q^{\rm kin} \sim \Nc  n_B^{5/3}/m_Q$. 
That is,
\beq
\calP_Q^{\rm kin} \simeq \Nc^2 \calP_N^{\rm kin} \,,
\eeq
meaning that the kinetic pressure of quark matter is about $\Nc^2 \sim 10$ times bigger than that of nuclear matter. 
Even before including interactions, the quark matter description can give large pressure.
This implies that quark matter descriptions are good starting points to construct stiff equations of state \cite{Annala:2019puf}.

Having seen that the non-relativistic quark matter already has the kinetic pressure,
we now quantify a simple parametric relation between the relativistic limit of quark matter and the mass-radius relation of a quark star.
The equation of state is
\beq
\varepsilon_Q^{\rm rela} = c n_B^{4/3} + \calB\,,~~~~~~
\calP_Q^{\rm rela} = c \frac{\, n_B^{4/3} \,}{3} -\calB
= \frac{\, \varepsilon_Q^{\rm rela} - 4 \calB \,}{3} \,,~~~~~~
c_s^2 = \frac{1}{\, 3 \,} \,.
\eeq
where we have introduced a bag constant $\calB$ which is related to the onset density of quark matter, $\calP_Q^{\rm rela} (n_B^{\rm onset}) = 0$.
There is a convenient expression relating the value of $\calB$ to the maximum mass $M_{\rm max}$ and the corresponding radius $R|_{ M_{\rm max} }$ \cite{Witten:1984rs}
\begin{eqnarray}
M_{\rm max}/M_{\odot} = 2.03 M_\odot \bigg( \frac{\, 56\,{\rm MeV/ fm^{3}} \,}{ \mathcal{B} } \bigg)^{1/2} \,,
~~~~~
R |_{ M_{\rm max} } = 10.7\,{\rm km} \times \frac{M_{\rm max} }{\, 2.0 M_\odot \,} \,.
\end{eqnarray}
The $2M_\odot$ constraint can be satisfied for $\mathcal{B} \lesssim 56\, {\rm MeV/fm^3}$. 
This expression suggests that the relativistic kinetic energy alone can produce stiff equations of state for a small bag constant.
In hybrid models with first order phase transitions, $\calB$ is taken large enough
to make phase transitions from nuclear to quark matter possible.
If we do not stick to models with such phase transitions, however, it is not necessary to choose a large $\calB$ \cite{Alford:2004pf}.
Although the choice of the bag constant is an important issue to be clarified,
there is a good chance to construct stiff equations of state based on quarks.

Finally, we briefly mention the possible structure of $\varepsilon_Q^{\rm int}$.
At very high density $n_B \gg \lqcd^3$, one should not use contact type interactions but directly use interactions mediated by gluons.
Interaction terms like $\sim n_B^2/\lqcd^2$ or $\sim n_B^3/\lqcd^5$, the contact many-body repulsions similar to those used in nucleonic models,
are omitted from the candidates to stiffen equations of state.
Instead, there are alternative candidates to stiffen equations of state, which can be attractive.
For instance, $\varepsilon_Q^{\rm int} = - b n_B^{2/3}$ with $b>0$, which comes from attractive correlations near the quark Fermi surface,
can stiffen equations of state \cite{Kojo:2014rca},
\beq
\calP^{\rm int}_Q = \frac{\,b \,}{\, 3 \,}  n_B^{2/3} ~~(>0) 
\,.
\eeq
The natural candidates for such interaction terms are pairing correlations near the Fermi surface, e.g., diquarks or baryonic correlations.
Unlike the interactions with higher powers, the high density limit leads to the equations of state dominated by the kinetic energy,
i.e., the conformal limit $\calP \rightarrow \varepsilon/3$.

\section{Quark saturation and IdylliQ model} \label{sec:q-sat}

The IdylliQ model is an ideal model of quarkyonic matter \cite{Fujimoto:2023mzy}.
It contains the concept of duality which allows us to describe dense matter using two languages,
baryonic and quark descriptions.
This model is particularly useful to describe the domain intermediate between nuclear and quark matter
where the effective degrees of freedom are not clear-cut.
By describing many-body states using both baryonic and quark languages,
one can cover the same physics from multiple angles.

\subsection{Sum rules} \label{sec:sum_rules}

In general, quantum many-body systems are very complicated.
We compromise by adopting a coarse-grained descriptions.
We characterize the state of matter by describing which states of quarks and baryons are occupied \cite{Kojo:2025vcq}.

As a preparation, let us begin by considering quarks in a single baryon.
We denote $\varphi (\vq-\vk/\Nc)$ as a quark distribution within a baryon with the momentum $\vk$.
The normalization condition is $\int_{\vq} \varphi (\vq) = 1$.
Assuming $\varphi (\vq)$ to be a function of $\vq^2$,
the average and variance of quark momentum in a baryon are characterized as
[$\int_{\vq} \equiv \int \rmd \vq/(2\pi)^3$]
\beq
\la \vq \ra_{\vk}
= \int_{\vq} \vq \varphi (\vq-\vk/\Nc)
= \vk/\Nc \,,~~~~~~
\la \vq^2 \ra_{\vk=0}
= \int_{\vq} \vq^2 \varphi (\vq)
\simeq \lqcd^2
\,.
\eeq
The average momentum of a quark is small but its variance is substantial.
This means that quarks in a baryon are very energetic.
At a given $\vk$, the average energy of a quark is
\beq
\la E_Q  \ra_{\vk}
= \int_{\vq} E_Q ( \vq + \vk/\Nc ) f_Q (\vq)
\simeq 
\la E_Q \ra_{\vk=0}
+ \frac{\, k_i k_j \,}{\, 2! \Nc^2 \,} \bigg\la \frac{\, \partial^2 E_Q \,}{\, \partial q_i \partial q_j \,} \bigg\ra_{\vk=0} + \dots \,.
\eeq
Summing $\Nc$ quark contributions, we obtain an energy expression similar to
the baryon mass term plus the non-relativistic kinetic term:
$\Nc \la E_Q \ra_{\vk} \sim E_B(\vk) \simeq M_B + \vk^2/2M_B + \dots$.
We note that quarks are energetic and the mechanical pressure is large within the baryon,
but what contributes to the thermodynamic pressure is the baryon kinetic energy of $\sim 1/\Nc$ which is suppressed by the large baryon mass.

Using the quark momentum distribution $\varphi$,
now we consider the following sum rule
in which we integrate quark contributions from each baryon to get the quark distribution in a many-baryon system \cite{Kojo:2025vcq},
\beq
f_Q (\vq;n_B) = \int_{\vk} f_B (\vk;n_B) \varphi (\vq-\vk/\Nc) \,,
\label{eq:fQ}
\eeq
where $f_B(\vk)$ and $f_Q(\vq)$ are occupation probabilities of baryonic and quark states, respectively.
Here $f_Q$ represents the distribution for a given color,
$f_Q \equiv f_Q^R = f_Q^G = f_Q^B$. 
Integrating $f_Q$ over $\vq$, one finds
\beq
n_B 
= \int_{\vq} f_Q (\vq;n_B) = \int_{\vk} f_B (\vk;n_B) 
\,.
\eeq
This reflects the fact that a single baryon contains a single quark of a given color.

In dilute regime, one can derive approximate formulae for $f_Q$.
If $f_B (\vk)$ has the distribution which is substantial only for $|\vk| < \lqcd$,
we can regard $\vk/\Nc$ as small
and apply the factorization
\beq
f_Q (\vq;n_B) \simeq \varphi (\vq) \int_{\vk} f_B (\vk;n_B) = n_B \varphi (\vq) ~\sim~ O(n_B/\lqcd^3) \,.
\label{eq:scaling}
\eeq
Quark states are occupied only partially for $n_B \ll \lqcd^3$
but begin to violate the upper bound at large $n_B$.
We call the saturation of quark states simply {\it quark saturation} \cite{Kojo:2025vcq}.
The quark states are supposed to get saturated first at $\vq=0$;
the quark saturation density is given through
\beq
 f_Q (\vq = 0;n_B^{\rm sat}) = \int_{\vk} f_B (\vk;n_B^{\rm sat}) \varphi (\vk/\Nc) \,.
\eeq
Beyond $n_B^{\rm sat}$, the scaling in Eq.~\eqref{eq:scaling} violates the quark Pauli principle.
This suggests that the assumption used for the factorization, i.e., dropping $\vk/\Nc$, must be invalid at $n_B \ge n_B^{\rm sat}$.

\subsection{ Quark Saturation and Rapid Stiffening of the Equation of State } \label{sec:quark_sat}

How can we avoid the violation of the quark Pauli principle?
The first possibility is that  the sum rule becomes invalid.
This may occur because baryons change their structure.
Another possibility is that baryons are not radically modified but occupy states in an unusual way.

Below we focus on the second possibility assuming the sum rule to be valid.
To illustrate the idea, we change the variable in the sum rule as $\vk' = \Nc \vk$,
\beq
f_Q (\vq;n_B) = \Nc^3 \int_{\vk'} f_B ( \Nc \vk'; n_B) \varphi (\vq-\vk') \,,
~~~~~~~ n_B = \Nc^3 \int_{\vk'} f_B ( \Nc \vk'; n_B) \,.
\eeq
Note that there is an overall factor $\Nc^3$ which would violate the condition $f_Q \le 1$.
In the dilute regime the violation does not occur since the range of integration is limited as $ |\vk'| \le k_F^B/\Nc$ and the factor $\Nc^3$ can be cancelled.
Due to this restriction on $\vk'$, one can use the factorization to reproduce the previous conclusion.
But beyond $n_B^{\rm sat}$, the factorization must be violated, as we mentioned before.
This means that there should be a contribution of $\vk' \sim O(\Nc^0)$,
but apparently this would conflict with the baryon number soon after the saturation, $n_B \sim \lqcd^3$, not $\sim \Nc^3 \lqcd^3$.

The solution to this apparent paradox is to consider $f_B \sim 1/\Nc^3$ for $|\vk'|$ in the domain $[0, \sim \lqcd]$
or equivalently for $|\vk|$ in the domain of $[0, \sim \Nc \lqcd]$.
For the bulk part of $f_B$ the states are under occupied.
Meanwhile, at high momentum of $\vk \sim \Nc \lqcd$, baryons become less influential to quark states at low momenta since the phase space decouples.
For such domain $f_B (\vk \sim \Nc \lqcd)$ can be $\sim 1$ but its thickness $\Delta k$ must be limited to $\Delta k \sim 1/\Nc^2$ or $\Delta k' \sim 1/\Nc^3$
to cancel the factor $\Nc^2$ arising in the phase space factor $4\pi k^2 \rmd k \sim (\Nc \lqcd)^2 \rmd k$.

The above description leads to the picture that
the quark Fermi sea begins to form with the quark saturation and eventually evolve into the usual quark Fermi sea.
The resulting equations of state are those of quarks
which can be stiff as we have discussed in Sec.~\ref{sec:quark}.
In terms of baryonic descriptions, the equations of state are also stiff as baryons at low momenta, which have large mass energy but little pressure,
are quenched; they are forced to occupy the high momentum domain of $\Nc \lqcd$,
naturally leading to pressure of $\sim \Nc \lqcd^4$, rather than $\sim \lqcd^4/\Nc$ expected from non-relativistic baryons.
The above-mentioned momentum shell structure in baryon distributions
and associated stiffening was first proposed by McLerran and Reddy \cite{McLerran:2018hbz}.

\subsection{ IdylliQ model } \label{sec:IdylliQ}

Below we consider an ideal model to explicitly demonstrate the picture presented in the last section.
An ideal model of quarkyonic matter, the {\it IdylliQ} model, is constructed based on the following assumptions:
\\
(i) The model neglects all interactions except those responsible for quark confinement; 
the effect of confinement is implemented by requiring that quarks exist only as constituents of baryons.
Under this assumption, the energy density is given by 
\beq
\varepsilon [f_B] = \int_k E_B (k) f_B (k) \,.
\label{eq:en_fB}
\eeq
For the moment we ignore the spin and flavor factors.
\\
(ii) We ignore the density dependence of the internal quark momentum distribution $\varphi$.
In realistic settings, the width of $\varphi$ in momentum space is expected to shrink as baryons swell in dense matter. 
However, to retain tractability, we fix $\varphi$ from low to high density.
\\
(iii) We adopt a specific functional form for $\varphi$ that allows for an analytic inversion of the sum rule, enabling us to express 
$f_B$ as a functional of the quark distribution $f_Q$.
Explicitly, we take:
\beq
\varphi (q) = \frac{\, 2 \pi^2 \,}{\, \Lambda^3 \,} \frac{\, e^{-q/\Lambda} \,}{\, q/\Lambda \,} \,,
\eeq
which is the inverse of the operator
\beq
L = - \nabla_q^2 + \frac{1}{\, \Lambda^2 \,} \,,~~~~~ L \big[ \varphi ({\bf q}-{\bf p} ) \big] = (2\pi)^3 \delta (\bf{q} - \bf{p}) \,.
\eeq
With this we can express $f_B$ as
\beq
f_B (N_c q) = \frac{\, \Lambda^2 \,}{\, N_c^3 \,} L \big[ f_Q (q) \big] \,.
\label{eq:fB_from_fQ}
\eeq
For instance, in a domain where quark states are saturated, i.e., $f_Q (q) = 1$, then
the derivative term vanishes and we obtain $f_B (N_c q) = 1/N_c^3$.

Now the rest is straightforward; we minimize $\varepsilon$ by optimizing $f_B$ at each momentum.
To perform the minimization subject to a fixed baryon density $n_B$, 
we must impose this constraint during the variation of $f_B (k)$.
A practical way to enforce this is to consider paired variations in momentum space such that
\begin{eqnarray}
\delta f_B ({\bf k}_1) + \delta f_B ({\bf k}_2) = 0 \,.
\end{eqnarray}
This condition ensures that the total baryon density remains unchanged during the variation. 
The corresponding change in the energy density is then given by
\begin{align}
\delta \varepsilon 
= E_B (k_1) \delta f_B ({\bf k}_1) + E_B (k_2) \delta f_B ({\bf k}_2) 
= \big[\, E_B (k_1) - E_B (k_2) \, \big] \delta f_B ({\bf k}_1) \,. 
\end{align}
This condition implies that relocating a particle from ${\bf k}_2$ to ${\bf k}_1$ reduces the total energy if $|{\bf k}_2| > |{\bf k}_1|$.
Consequently, the optimal distribution $f_B (k)$ must concentrate particles at low momenta.
Defining $k_{\rm sh}$ as the largest momentum below which $f_B$ is nonzero,
we conclude that $f_B (k)$ vanishes for $k > k_{\rm sh}$,
while it is maximized for $k \le k_{\rm sh}$.

The maximum value that $f_B$ can attain is governed by the sum rule constraint.
If no quark momentum states are saturated, then $f_B$ can reach the maximum value of 1, 
recovering the standard ideal gas distribution, $f_B^{\rm ideal} (k) = \Theta (k_{\rm sh}-k)$.
However, when some domain in quark momentum space is saturated, 
the sum rule enforces a bound on $f_B$, limiting it to the maximal value $1/N_c^3$.
Assembling these elements, the baryon distribution takes the form
\beq
f_B (k) = \frac{1}{\, \Nc^3 \,} \Theta (k_{\rm bu} - k) + \Theta( k_{\rm sh} - k ) \Theta (k -k_{\rm bu}) \,.
\label{eq:sol_fB}
\eeq
The bulk domain $k \le k_{\rm bu}$ corresponds to the domain where quarks are saturated 
and hence $f_B$ reaches the maximum value $1/\Nc^3$.
For the unsaturated domain for $k \ge k_{\rm bu}$, $f_B$ reaches the maximum value $1$.  
At large density with $k_{\rm sh} \sim \Nc \Lambda$ or beyond, the interval $k_{\rm sh} - k_{\rm bu}$ to be the order of $\sim \Lambda^2/\Nc^2$.
This can be seen from the baryon shell contributions to $f_Q$ at $q=0$,
\beq
f_Q^{\rm from-shell} (0) 
\sim \int_{k_{\rm bu}}^{k_{\rm sh}} \rmd k k^2 \varphi (\vk/\Nc)
\sim ( k_{\rm sh} - k_{\rm bu} ) \big( \Nc \Lambda \big)^2 \varphi (\Lambda) \,.
\eeq
We need $k_{\rm sh} - k_{\rm bu} \sim \Lambda/\Nc^2$ to cancel the factor $\Nc^2$ arising from the second factor. 

The corresponding quark distribution takes the form
\beq
f_Q (q) = \Theta (q_{\rm bu} - q) + f_Q^{f_B=1} (q) \Theta( q_{\rm sh} - q ) \Theta (q -q_{\rm bu}) + f_Q^{f_B=0} (q) \Theta( q-q_{\rm sh} )\,,
\eeq
where $k_{\rm bu} = \Nc q_{\rm bu}$ and $k_{\rm sh} = \Nc q_{\rm sh}$, required from Eq.~\eqref{eq:fB_from_fQ}.
The functions $ f_Q^{f_B=1}$ and $ f_Q^{f_B=0}$ satisfy
\beq
1 = \frac{\, \Lambda^2 \,}{\, N_c^3 \,} L \big[ f_Q^{f_B=1} \big] \,,~~~~~~~
0 = \frac{\, \Lambda^2 \,}{\, N_c^3 \,} L \big[ f_Q^{f_B=0} \big] \,.
\eeq
To find the solutions, it is convenient to notice that the operator $L$ contains two derivatives and a constant.
The homogeneous solutions satisfying $L[y_\pm]=0$ are
\beq
y_\pm (q) = \frac{\, \rme^{\pm q/\Lambda} \,}{\, q/\Lambda\,} \,.
\eeq
The solutions for $ f_Q^{f_B=1}$ and $ f_Q^{f_B=0}$ take the form
\beq
f_Q^{f_B=1} (q) = \Nc^3 + c_+ y_+(q) + c_- y_- (q) \,,~~~~~~~
f_Q^{f_B=0} (q) = d_- y_- (q) \,.
\eeq
We note that, to satisfy $0 \le f_Q^{f_B=1} \le 1$, we need to arrange $c_+$ and $c_-$ carefully to cancel the large constant $\Nc^3$.
This implies that a baryon is a consequence of sophisticated cancellations of different quark contributions.
We also note that the interval between $q_{\rm bu}$ and $q_{\rm sh}$ is very small.
Since the function $y_+$ is a growing function while $y_-$ is a decreasing function in $q$,
the cancellation among three terms in $f_Q^{f_B=1}$ cannot hold for a very long interval in $q$.
This is consistent with the observation of $k_{\rm sh} - k_{\rm bu} \sim \Lambda/\Nc^2$;
this leads to $q_{\rm sh} - q_{\rm bu} \sim \Lambda/\Nc$.

There are at least two practical ways to fix $k_{\rm bu}$ and $k_{\rm sh}$ as a function of $n_B$.
The first one is to focus on the boundary conditions at $q_{\rm bu}$ and $q_{\rm sh}$, matching the functions smoothly across the boundaries. 
This approach utilizes the global constraints on the total baryon number to uniquely determine these parameters.
Operating $L$ on the step functions yield the delta- and the derivative of the delta-functions,
which would violate the condition $0 \le f_B \le 1$.
In order to avoid this, the coefficients of these singular functions must vanish.
Then we obtain a set of conditions
\begin{align}
1 = f_Q^{f_B=1} (q_{\rm bu}) \,,~~~~
f_Q^{f_B=1} (q_{\rm sh}) = f_Q^{f_B=0} (q_{\rm sh}) \,,~~~~
0 = \frac{\, \rmd f_Q^{f_B=1} \,}{\, \rmd q \,} \bigg|_{q_{\rm bu}}\,,~~~~
\frac{\, \rmd f_Q^{f_B=1} \,}{\, \rmd q \,} \bigg|_{q_{\rm sh}}
 = \frac{\, \rmd f_Q^{f_B=0} \,}{\, \rmd q \,} \bigg|_{q_{\rm sh}}
 \,.
\end{align}
We have five unknown $(q_{\rm bu}, q_{\rm sh}, c_+, c_-, d_-)$
while four boundary conditions plus one condition $n_B = \int_{\vk} f_B (\vk) $ are available. 
Hence unknown constants can be determined uniquely.

Another way to determine constants is to first fix $k_{\rm bu}$ and $k_{\rm sh}$
and then compute $(c_+, c_-, d_-)$ as outputs.
First we define
\beq
f_Q^{ [0,k_s] } (q) \equiv \int_{\vk} \Theta( k_s - k ) \varphi (\vq - \vk/\Nc ) \,,~~~~~~~
f_Q^{ [k_1,k_2] } (q) \equiv f_Q^{ [0,k_2] } (q) - f_Q^{ [0,k_1] } (q) \,.
\eeq
The computation yields three functions of $q$ with the coefficients as functions of $k_s = \Nc q_s$,
\beq
f_Q^{ [0,k_s] } (q)
= \Nc^3 \Theta (q_s -q)
+  A_{\rm in}^{ [0,k_s] } \Theta (q_s-q) y_s (q)
+  A_{\rm out}^{ [0,k_s] } \Theta (q-q_s) y_- (q) \,,
\eeq
where the coefficients are
\beq
A_{\rm in}^{ [0,k_s] } = - (1+q_s/\Lambda ) \rme^{-q_s/\Lambda} \,,~~~~~~~
A_{\rm out}^{ [0,k_s] } = (q_s/\Lambda ) \cosh(q_s/\Lambda) - \sinh( q_s/\Lambda) \,.
\eeq
The $f_B$ in Eq.~\eqref{eq:sol_fB} leads to
\begin{align}
f_Q(q) 
&= \frac{1}{\, \Nc^3 \,} f_Q^{ [0,k_{\rm bu}] } (q) + f_Q^{ [k_{\rm bu}, k_{\rm sh}] } (q)
\notag \\
& = \Theta (q_{\rm bu} - q ) 
+ y_s (q) \bigg[ \bigg( \frac{1}{\, \Nc^3 \,} - 1 \bigg) A_{\rm in}^{ [0,k_{\rm bu} ] } \Theta (q_{\rm bu}-q)
+  A_{\rm in}^{ [0,k_{\rm sh}] } \Theta (q_{\rm sh} -q) 
\bigg]
\notag \\
& \hspace{2.1cm}
+ y_- (q) \bigg[ \bigg( \frac{1}{\, \Nc^3 \,} - 1 \bigg) A_{\rm out}^{ [0,k_{\rm bu} ] } \Theta (q - q_{\rm bu})
+  A_{\rm out}^{ [0,k_{\rm sh}] } \Theta ( q-q_{\rm sh} ) 
\bigg] \,.
\end{align}
This expression can be rearranged into
\begin{align}
f_Q(q) 
& = \Theta (q_{\rm bu} - q ) 
+ \Theta (q_{\rm bu}-q) y_s (q) 
\bigg[ \bigg( \frac{1}{\, \Nc^3 \,} - 1 \bigg) A_{\rm in}^{ [0,k_{\rm bu} ] } 
+ A_{\rm in}^{ [0,k_{\rm sh}] }
\bigg]
\notag \\
&~~
+ \Theta (q_{\rm sh} -q) \Theta( q - q_{\rm bu} )
\bigg[
  y_s (q) A_{\rm in}^{ [0,k_{\rm sh}] } 
  + y_- (q) \bigg( \frac{1}{\, \Nc^3 \,} - 1 \bigg) A_{\rm out}^{ [0,k_{\rm bu} ] } 
\bigg] 
\notag \\
&~~
+ \Theta ( q-q_{\rm sh} ) y_- (q) 
\bigg[ \bigg( \frac{1}{\, \Nc^3 \,} - 1 \bigg) A_{\rm out}^{ [0,k_{\rm bu} ] } 
+  A_{\rm out}^{ [0,k_{\rm sh}] } 
\bigg] 
 \,.
\end{align}
The saturation condition for $0 \le q \le q_{\rm bu}$ requires the coefficient of $y_s (q) \Theta ( q_{\rm bu} - q )$ to vanish.
This leads to an equation relating $q_{\rm sh}$ and $q_{\rm bu}$.
Then, the condition $n_B = \int_{\vk} f_B (\vk)$ gives a condition to determine either of $q_{\rm sh}$ or $q_{\rm bu}$.
After determining $q_{\rm sh}$ and $q_{\rm bu}$, we can fix the other parameters $(c_+,c_-,d_-)$ as functions of $(q_{\rm bu}, q_{\rm sh})$.

\section{ Mitigating the hyperon puzzle in neutron stars } \label{sec:hyperons}

In this section, we extend the IdylliQ model to multi-flavor environments to address one of the most persistent challenges in neutron star physics: 
the hyperon puzzle \cite{Lonardoni:2014bwa,Tolos:2020aln}.
The emergence of hyperons softens equations of state
by converting energetic neutrons into non-relativistic hyperons.
This occurs when the Fermi energy of neutrons exceed the mass of (in-medium) hyperons,
$E_F^{n} \ge m^*_Y$.
This conversion makes the equation of state difficult to pass the $2M_\odot$ constraints.

In typical models, the conversion $n \rightarrow Y$ occurs around $2$–$3n_0$.
One way to avoid this problem is to introduce strong short-range repulsion between hyperons and nucleons \cite{Gerstung:2020ktv}.
The two-body repulsion $YN$ is usually not sufficient, and three-body $YNN$ repulsion is added to increase the in-medium masses of hyperons \cite{Lonardoni:2014bwa}. 
However, scenarios based on many-body repulsion raise questions regarding the convergence of the many-body expansion. 
Another problem is that, as in usual arguments based on many-body repulsion, $R_{2.1}$ becomes considerably smaller than $R_{1.4}$, $R_{2.1} - R_{1.4} \sim -1$ km.
To avoid these issues, discussions at a more fundamental level are needed.

Below we examine a scenario based on quark saturation.
This yields statistical $YN$ ($YNN$) repulsions.
There are several advantages in this strategy:
(i) it is based on the quark Pauli blocking which is inevitable at large density;
(ii) the statistical repulsion becomes stronger at higher density;
and (iii) there is no worry about double counting of quarks; we manifestly treat quarks from each baryon.

We extend the IdylliQ model to the multi-flavor case \cite{Fujimoto:2024doc}. The extension is straightforward. 
The sum rules become
\beq
f_Q (\vq) = \sum_{B=p,n,\Lambda,..} \int_{\vk} f_B (\vk) \varphi^B_Q (\vq-\vk/\Nc) \,,~~~~~(Q=u,d,s) \,,
\eeq
where $\varphi^B_Q$ is the quark distribution for a given baryon $B$.
That is, we simply add all possible baryon species and collect quark contributions from them. 
It is clear that once the quark saturation sets in, 
there appears the statistical correlation among baryon species.
For simplicity we assume the spatial distributions are common for different baryon and quark species,
\beq
\varphi^B_Q (\vq) = N^B_Q \varphi (\vq) \,, 
\eeq
where $N^B_Q$ is the quark fraction $Q=u,d,s$ for a given baryon $B = p,n,\Lambda,\cdots$.
For example, $N^p_u = 2/3$, $N^p_d = 1/3$, $N^p_s = 0$, and so on.

For concreteness, we consider a charge neutral matter made of $n(udd)$ and $\Lambda_0(uds)$.
Let us assume that the neutral density is high and $d$ quark states are saturated before the emergence of $\Lambda_0$.
(Whether $\Lambda_0$ appears earlier or not is not so crucial, since the $d$-quark saturation occurs soon after the appearance of $\Lambda_0$.)
We also neglect $n$-$\Lambda_0$ interactions to emphasize the statistical aspects.

\begin{figure}[t]
\vspace{-1.5cm}
\centering
\includegraphics[width=.7\textwidth]{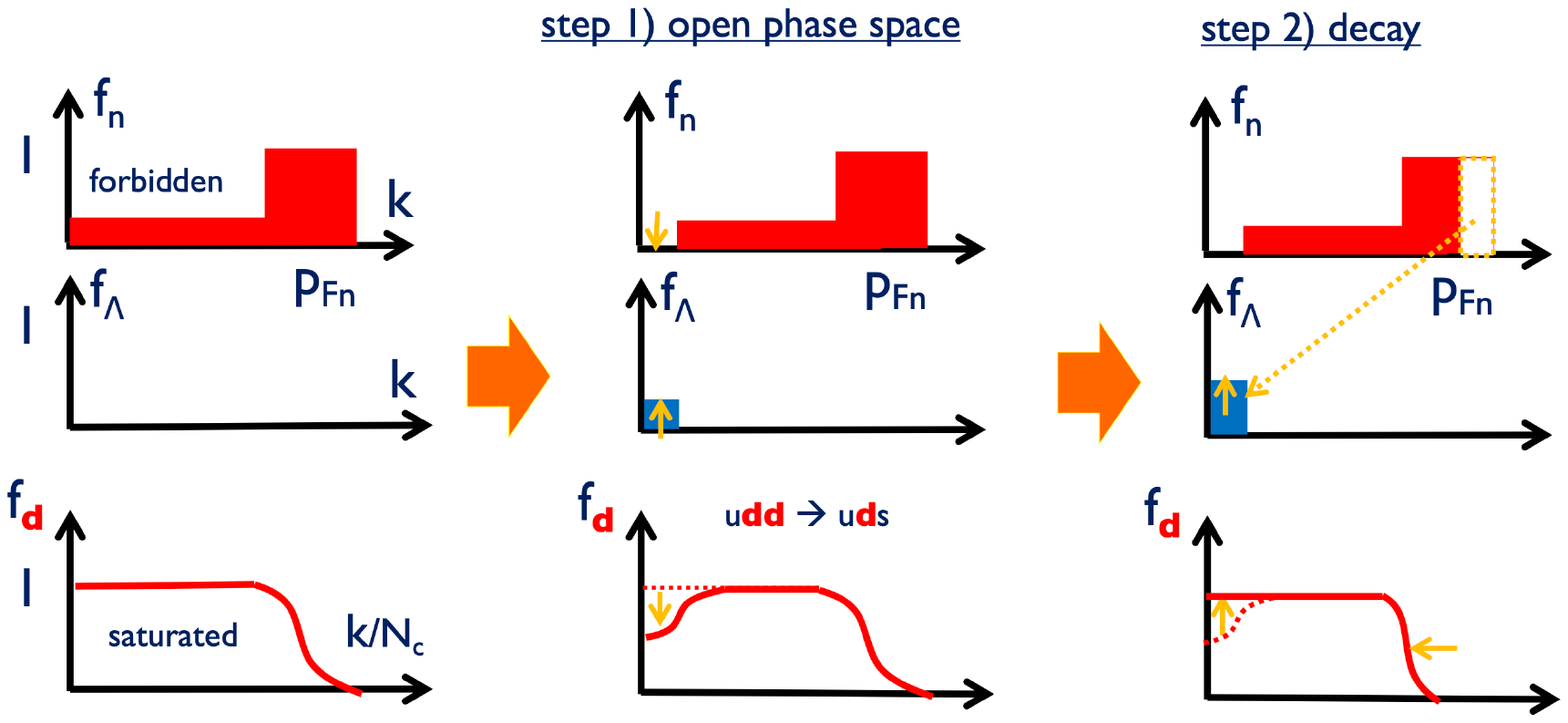}
\vspace{-1.5cm}
\caption{ The statistical repulsion between the saturated $d$-quark Fermi sea and hyperons.
The necessity to open the phase space for $d$-quarks requires extra energetic cost of $m_\Lambda - m_n \simeq M_s - M_{u,d}$.
}
\label{fig:hyperon}
\end{figure}

We consider the process $n(k_F^n) \rightarrow \Lambda (k=0)$,
where a neutron with the Fermi energy $E_F^n = E_n (k_F^n)$ decays to $\Lambda_0$ at rest (Fig.~\ref{fig:hyperon}).
In purely hadronic model, this decay occurs when $E_F^n \ge m_{\Lambda_0}$.
But this process is forbidden in the presence of the $d$-quark saturation.
The $d$-quarks at low momenta are saturated by neutrons in the bulk domain,
and having $\Lambda_0 (uds)$ at low momenta adds extra $d$-quark with the low momentum, violating the quark Pauli principle.

In the IdylliQ model, $d$-quark saturation imposes the condition for baryons at a momentum $\vk$ as \cite{Fujimoto:2024doc}
\beq
\frac{ 1 }{\, \Nc^3 \,} = \frac{2}{\, 3 \,} f_{n} (\vk) + \frac{1}{\, 3 \,} f_{\Lambda_0} (\vk) \,,~~~~~~~~~ 0 \le |\vk| \le \Nc q^d_{\rm bu} \,,
\eeq
where $d$-quark states are saturated to the momentum $q^d_{\rm bu} $.
The composition of $f_{n} (\vk)$ and $ f_{\Lambda_0} (\vk)$ must be determined by energetic arguments.
Note that the factor $2/3$ ($1/3$) appears in front of $f_n$ ($f_{\Lambda_0}$) 
because a neutron contain two (one) $d$-quarks.

Therefore, for $\Lambda_0$ to appear at rest, we first open the phase space for $d$-quarks at zero momentum.
This can be done by replacing $n(k=0)$ with $\Lambda_0 (k=0)$.
A neutron contains more $d$-quarks than $\Lambda_0$,
so the replacement opens the phase space for $d$-quarks.
Although this replacement costs the extra energy $m_\Lambda - m_n$, 
this also allows the decay process $n(k_F^n) \rightarrow \Lambda (k=0)$ which reduces the energy of the system.
These competing effects determine the onset of $\Lambda_0$.
The condition for the conversion $n \rightarrow \Lambda_0$ is
\beq
E_n (k_F^n) - m_{\Lambda} \ge m_{\Lambda} - m_n 
~ \rightarrow ~
\mu_B = E_n (k_F^n) \ge 2 m_{\Lambda} - m_n \,.
\eeq
In addition to usual chemical potential of $m_\Lambda$, 
$d$-quark saturation requires the extra chemical potential of $m_\Lambda - m_n \simeq 200$ MeV.

For reasonable model parameters leading to $d$-quark saturation at $\sim 2n_0$,
the onset of $\Lambda_0$ shifts from $\sim 2n_0$ to $\sim 5n_0$.
Moreover, even after the appearance of $\Lambda_0$,
the equations of state do not soften substantially because
the occupation probability of $\Lambda_0$ in the bulk is $\sim 1/\Nc^3$.
All these effects mitigate the hyperon puzzle.

We note that $d$-quark saturation does not suppress the appearance of, e.g., $\Sigma_+ (uus)$ or $\Xi_0 (uss)$ hyperons,
since they do not have $d$-quarks.
The former is usually not considered because it is a positively charged particle;
in neutron star setup the charge chemical potential is negative which disfavors $\Sigma_+$;
if the charge chemical potential were positive, protons should appear first, saturate $u$-quark states, disfavoring $\Sigma_+$.
As for $\Xi_0$, it contains two strange quarks and hence require the extra chemical potential of $\sim 200$ MeV.
Then, the onset density is $n_B \sim 5n_0$ as in the case of $\Lambda_0$.

\section{Hints of quarkyonic matter from QCD-like theories} \label{sec:hints}

Finally, we mention testbeds which can test the concepts presented in this article.

Two-color QCD with two-flavor quarks
is a QCD-like theory whose finite-density systems can be simulated by the lattice Monte-Carlo simulations \cite{Kogut:1999iv}.
In this theory baryons are diquarks which are bosonic and hence the low density behaviors are different from QCD,
but, at very high density, the quark matter is formed anyway and one can learn some aspects of dense matter.
The phase structure was studied in early studies \cite{Sun:2007fc,Brauner:2009gu},
but more detailed characterization for equations of state was done only recently \cite{Kojo:2021hqh}.
In particular, the sound speed peak was discovered in the hadron-to-quark transition domain \cite{Iida:2022hyy}.
Similar trends were also found in QCD at finite isospin density \cite{Brandt:2022hwy,Abbott:2024vhj}.
The lattice equations of state in Ref.~\cite{Abbott:2024vhj}, which cover from hadronic to the perturbative QCD domain,
agree with the predictions of quark-meson-type models \cite{Chiba:2023ftg,Kojo:2024sca,Brandt:2025tkg}
in which the effective degrees of freedom smoothly change from hadrons to quarks;
the identification of degrees of freedom is the key for such agreement.
Further, hyperon puzzle can be also studied in 2+2 flavor setup,
and model studies found statistical repulsion \cite{Nagatsuka:2025ijc}; this setup can be also studied on the lattice.

In two-color QCD, not only bulk properties but also more microscopic aspects have been also studied.
One of important microscopic properties is the presence of energy gaps associated with the diquark condensates.
Unlike in three-color QCD, the diquarks are color-singlet
so that neither color-electric nor magnetic gluons are screened in the static limit \cite{Rischke:2000cn}.
This suggests that the gluons remain confining,
and this picture is consistent with the lattice computations for the Polyakov loop \cite{Boz:2019enj}, topological susceptibility \cite{Iida:2019rah,Iida:2024irv}, 
and gluon propagators \cite{Boz:2018crd}.
These results seem consistent with studies based on effective models with the diquark gap $\Delta$ of $\sim 200$ MeV
and Polyakov loops \cite{Kurebayashi:2025fcw}.
For weak coupling estimates of the gap and its density evolution, see, e.g., Refs.~\cite{Fukushima:2024gmp,Fujimoto:2024pcd}.

Another important theory to study quarkyonic matter is QCD in spatially one dimension.
The theory is confining as the electric flux must be string-like because there is only one available direction.
The strength of the screening does not change from vacuum to finite density
as the phase space for color-screening remains the same at any densities \cite{Kojo:2011fh}.
Recently this system was studied using the Hamiltonian-based simulations \cite{Hayata:2023pkw,Fujikura:2026tqk}.
Not only equations of state but also the quark occupation probabilities were computed.
The simulations found that the evolution of the quark momentum distribution
can be interpreted as the quark Fermi sea plus the diffused Fermi surface.

\section{Summary} \label{sec:summary}

Quarkyonic matter is a good candidate to describe a smooth transition from nuclear to quark matter.
Confinement persists during such a transition until gluons are strongly screened by medium effects.
The density interval for such transition is 2--$5n_0$, directly related to the structure of neutron stars
with the mass range of $1.4M_\odot$--$2.2M_\odot$.
With confining gluons, we consider quarks to be trapped within baryons,
nevertheless found that the quark Fermi sea is formed due to the saturation of quark states which begins from low momenta.
The density of the quark saturation is $\sim 2$--$3n_0$, about a half of the density for the baryon core to overlap.
The onset of the quark saturation signals the beginning of quark matter formation
with which $P(\varepsilon)$ is forced to approach the quark matter scaling.
The disparity between soft nuclear equations of state and stiff quark equations of state
requires the appearance of the sound speed peak.
The mechanism of quark saturation also mitigates the hyperon softening problem
by disfavoring hyperons containing $d$-quarks.

In this article we have not presented several important directions to extend the IdylliQ models.
One is the possible quark substructure effects on conventional nuclear physics \cite{Koch:2024qnz,McLerran:2024rvk}.
The other is the finite temperature extension \cite{Bluhm:2024uhj,Bluhm:2026hoj}.
The former would open a new avenue for the quark parton distribution such as a gravitational form factor.
The latter has direct impacts on heavy-ion collisions and neutron star mergers.
The quarkyonic matter conceptions would offer phenomenological effects which have not been explored in other descriptions.
Further studies are called for.

\section*{Acknowledgments}
I thank  the participants of the workshop, {\it From Quarks to Neutron Stars: Insights from kHz gravitational waves} held at University of Tokyo, for fruitful discussions.
This article is based on several collaborations.
The contents of the neutron star part are mainly based on collaboration with G. Baym and T. Hatsuda \cite{Kojo:2021wax}.
The contents about Quarkyonic matter 
are based on recent collaboration with Y. Fujimoto and L. McLerran \cite{Fujimoto:2023mzy,Fujimoto:2024doc}.
Several conceptions have been tested in QCD-like theories
worked out with 
R. Chiba \cite{Chiba:2023ftg,Chiba:2024cny,Kojo:2024sca}, Y. Kurebayashi \cite{Kurebayashi:2025fcw}, M. Nagatsuka \cite{Nagatsuka:2025ijc}, and D. Suenaga \cite{Kojo:2021hqh},
and the work with K. Iida, H. Liang, and H. Tajima \cite{Tajima:2024qzj}.

\section*{Funding}
This work was supported by JSPS KAKENHI Grant No. 23K03377 and 26K07077.

\bibliographystyle{iopart-num}

\bibliography{ref}

\end{document}